\documentclass[3p,times]{elsarticle}

\usepackage{ecrc}

\usepackage{amssymb}
\usepackage{amsthm}
\usepackage{amsmath}
\usepackage[utf8]{inputenc}
\usepackage{multirow}

\pdfmapfile{+txfonts.map}

\usepackage{xcolor}

\pdfmapfile{+txfonts.map}

\usepackage{epsfig}

\volume{00}

\firstpage{1}

\journalname{Physica A}

\runauth{Till Massing et al.}

\jid{procs}

\jnltitlelogo{Physica A}

\CopyrightLine{2020}{Published by Elsevier Ltd.}

\usepackage{amssymb}

\usepackage[figuresright]{rotating}

\begin{document}

\begin{frontmatter}







\title{On the parametric description of log-growth rates of
cities' sizes of four European countries and the USA}

 \author[label1]{Till Massing\fnref{label4}}
 \author[label3]{Miguel Puente-Ajov\'{\i}n\fnref{label4}}
  \author[label3]{Arturo Ramos\corref{cor1}}
 \address[label1]{Universit\"{a}t Duisburg-Essen,
 Fakult\"{a}t f\"{u}r Wirtschaftswissenschaften,
 Lehrstuhl für \"{O}konometrie, Essen, Germany}
\address[label3]{Departmento de An\'alisis Econ\'omico, Universidad de Zaragoza, Zaragoza, Spain}

\cortext[cor1]{Corresponding author\\
E-mail address: aramos@unizar.es}

\fntext[label4]{till.massing@uni-due.de (T. Massing), mpajovin@unizar.es (M. Puente-Ajov\'{\i}n)}

\begin{abstract}
We have studied the parametric description of the distribution of the log-growth rates of the sizes of cities of France, Germany, Italy, Spain and the USA. We have considered several parametric distributions well known in the literature as well as some others recently introduced.
There are some models that provide similar excellent performance, for all studied samples. The normal distribution is not the one observed empirically.
\end{abstract}

\begin{keyword} urban log-growth rate distribution \sep exponential distribution \sep exponential Generalized Beta 2 distribution \sep Student's $t$ distribution\sep mixture distributions




\end{keyword}

\end{frontmatter}

\section{Introduction}
\label{intro}

There has been a relatively recent interest in the parametric description of the log-growth rates of the sizes of firms and of
countries' outputs in its relation to Gibrat's law \citep{bottazzi2006edf,BotSec11,CanAmaLeeMeySta98,
CotFinHatArcBat19,CotHatArcBat17,Wit05,GatGuiGafGiuGalPal05,FagNapRov08,
FuPamBulRicMatYamSta05,
FujGuiAoyGalSou04,GalPal10,RicPamBulPonSta08,
Rozenfeld12022008,Sar19,
StaAmaBulHavLesMaaSalSta96}. In this strand of the literature, the asymmetric double Laplace
distribution is typically studied, instead of the normal distribution, to account for the observed exponential tails exhibited by the log-growth rates.
Later, in order to generalize the normal distribution at the same time as the asymmetric double Laplace distribution, there has also been considered the asymmetric Subbotin distribution.
It is generally accepted that these distributions provide a faithful description of the log-growth rates considered, above all the description of the approximately exponential tails.

Let us mention that in \cite{LiDonZhaWanWanDiSta17} it is
proposed a simple unified model to explain the growth dynamics of cities and scaling laws, where the model predicts that the size of cities grows linearly regardless of its current size.

For city sizes, the topic of growth rates has been studied, but mainly in a non-parametric way \citep{IoaOve03}, see also \cite{WuHe17} for a very recent article on this subject.
Perhaps because of the lack of good data sets until very recently, there have been very few studies that consider the parametric description of log-growth rates of all city sizes in a country. It is not until 2004 that Eeckhout considers the sample
of all USA Places in 1990 and 2000 (in the year 1990 without the Census Designated Places, CDPs) and takes the log-growth rates to be normally distributed, according to the strict fulfillment of Gibrat's law \citep{Gib31}. In this version of the law, the log-growth rates are i.i.d. normal variables that are added to yield the log-sizes, which at the end are described by a normal distribution (lognormal distribution for the sizes) since the convolution of normal distributions is exactly another normal distribution.

Very recently, there have appeared a few studies that deal with the topic of the parametric description of log-growth rates of city sizes
\citep{Ram17,SchTre13,SchTre16}. In the last two ones, the Student's $t$ distribution is proposed (for the German municipalities or \emph{Gemeinden}) as an appropriate model. In the other,
it has been tested whether the normal distribution provides a good description of the log-growth rates of USA city sizes, using different sources of data. The robust result that emerged is that the normal is rejected indeed by all of the statistical tests employed. Apart {}from that, other alternative distributions were studied: the asymmetric double Laplace normal (adLn)
of \cite{Ree02,Ree03,reed2004double} and above all, a new distribution called  \lq\lq double mixture exponential Generalized Beta of the second kind (dmeGB2)\rq\rq.
The latter provided the best fit, in terms of a number of statistical criteria, of all the proposals.

Likewise, the recent study \cite{PueRam15} deals with
data sets for the whole population of four major European countries: France, Germany, Italy and Spain. The computation of log-growth rates is relatively easy so natural questions
are whether the normal distribution, the adLn, the asymmetric Subbotin, or the dmeGB2, are a good description of these
log-growth rates. There are also newer models, introduced for the first time in this paper to study log-growth rates of city sizes, so that this paper makes the following contributions to the literature:
\begin{itemize}
\item[-] We introduce new models based on mixtures (as in the very recent paper \cite{KwoNad19}) of Student's $t$ distributions, improving in this way the approach of \cite{SchTre13,SchTre16}.
\item[-] The new models introduced are very competitive with respect to the dmeGB2, which is the best model of \cite{Ram17}.
\item[-] We observe that the tails of log-growth rates for the employed samples of those European countries and the USA are mainly non-exponential.
\item[-] We observe that the results do not depend on the type of urbanization (Europe vs the USA) or the age of cities (USA cities that are one, five, or nine decades old along the 20$^{\rm th}$ Century \citep{Cub11,GieSue14,SanGonVil14}).
\end{itemize}

The rest of the paper is organized as follows. In Section~\ref{meto} we will describe the methodology.
In Section~\ref{databases} we will account for the databases employed. In Section~\ref{results} we will present the empirical results. Finally we will offer some Conclusions.

\section{Method}\label{meto}

\subsection{Description of the distributions}

In this section we will introduce the distributions used throughout the paper for the (two consecutive periods) log-growth rates, denoted by
$$
g_{i,t}=\log x_{i,t}-\log x_{i,t-1}\in(-\infty,\infty)
$$
where $x_{i,t}$ is the population of city $i$ at time $t$. When a fixed $t$ is taken we will simply write $g\in(-\infty,\infty)$ for the variable of all log-growth rates of the cross-sections taken.

The first distribution we will consider is the well-known
normal distribution for the log-growth rates $g$. We thus set
$$
f_{{\rm n}}(g;\mu,\sigma)
=\frac{1}{\sqrt{2\pi}\sigma}
\exp\left(-\frac{(g-\mu)^{2}}{2\sigma^{2}}
\right)
$$
where $\mu$ is the mean of $g$ and $\sigma>0$ is its standard deviation according to this distribution.

The second model is the non-stan\-dar\-di\-zed Student's $t$ distribution for the log-growth rates $g$, see, e.g., \cite{JohKotBal95} and references therein, which is given by the probability density function
\begin{eqnarray}
f_{{\rm St}}(g;\mu,\sigma,\nu)=
\frac{\Gamma\left(\frac{\nu+1}{2}
\right)}{\Gamma\left(\frac{\nu}{2}\right)\sqrt{\pi \nu}\sigma}
\left(1+\frac{1}{\nu}\left(\frac{g-\mu}{\sigma}
\right)^2
\right)^{-\frac{\nu+1}{2}}\nonumber
\end{eqnarray}
where $\mu\in\mathbb{R}$ (location parameter), $\sigma>0$ (scale parameter), $\nu>0$ is the number of degrees of freedom, and $\Gamma(\cdot)$ denotes the Gamma function. Particular cases of this distribution are the Cauchy distribution ($\nu=1$)
and the normal distribution ($\nu=\infty$). If $1<\nu\leq 2$, the variance of the distribution becomes infinite. This distribution has been used to study the log-growth rates of the size of German cities \cite{SchTre13,SchTre16}.

The third distribution in our study will be
the asymmetric double Laplace normal distribution (adLn),
introduced by \cite{Ree02,Ree03,reed2004double} and later
used, e.g., by \cite{Man09}:
\begin{eqnarray}
& &f_{\rm adLn}(g;\alpha,\beta,\mu,\sigma)\nonumber\\
&&\quad=\frac{\alpha\beta}{2(\alpha+\beta)}{\rm e}^{-\alpha g}
\exp\left(\alpha\mu+\frac{\alpha^{2}\sigma^{2}}{2}\right)
\left(1+{\rm erf}\left(\frac{g-\mu-\alpha\sigma^{2}}
{\sqrt{2}\sigma}\right)\right) \nonumber\\
& &\quad\ -\frac{\alpha\beta}{2(\alpha+\beta)}
{\rm e}^{\beta g}
\exp\left(-\beta\mu+\frac{\beta^{2}\sigma^{2}}{2}\right)
\left({\rm erf}\left(\frac{g-\mu+\beta\sigma^{2}}{\sqrt{2}\sigma}\right)-1\right)\nonumber
\end{eqnarray}
where ${\rm erf}$ is the error function associated to the normal
distribution and $\mu\in\mathbb{R}$, $\alpha,\beta,\sigma>0$ are the four parameters of the distribution. It has the property that it approximates different exponential laws in each of its
two tails: $f_{\rm adLn}(g)\approx {\rm e}^{-\alpha g}$ when $g\to\infty$ and
$f_{\rm adLn}(g)\approx {\rm e}^{\beta g}$ when $g\to -\infty$. The body is approximately normal, although it is not
possible to exactly delineate the switch between the normal
and the exponential behaviours, since the adLn distribution
is the convolution of an asymmetric double Laplace with
a normal distribution.

We will describe here the asymmetric version \cite{BotSec11}
of a distribution initially introduced by \cite{Sub23} and later considered by \cite{bottazzi2006edf,CanAmaLeeMeySta98,GatGuiGafGiuGalPal05,
FagNapRov08,
FuPamBulRicMatYamSta05,FujGuiAoyGalSou04,
McD91,
RicPamBulPonSta08,Rozenfeld12022008,
StaAmaBulHavLesMaaSalSta96}. We will call it the asymmetric Subbotin distribution (aSub). The density function is then
$$
f_{{\rm aSub}}(g;a_l,a_r,b_l,b_r,\mu)=\left\{
\begin{array}{cc}
{\frac{1}{d}\exp\left(-\frac{1}{b_l}
\left|\frac{g-\mu}{a_l}\right|^{b_l}\right)} & {g\le\mu} \\
{\frac{1}{d}\exp\left(-\frac{1}{b_r}
\left|\frac{g-\mu}{a_r}\right|^{b_r}\right)} & {\mu\le g} \\
\end{array}
\right.
$$
where $d=a_l b_l^{1/{b_l}}\Gamma\left(1+1/{b_l}\right)+a_r b_r^{1/{b_r}}\Gamma\left(1+1/{b_r}\right)$. The quantities $a_l,a_r,b_l,b_r$ are positive shape parameters and $\mu$ is a location parameter that takes real values.
Particular cases of this distribution are the (asymmetric) double Laplace distribution when $b_l=b_r=1$, and the normal distribution when $a_l=a_r$ and $b_l=b_r=2$. The main interest of this distribution is to model the exponential tails of the observed log-growth rates (of the output of a country, or of firm sizes) and their (slight) deviations from this behaviour when the parameters $b_l,b_r$ depart {}from unity. Also, it is useful when studying the scaling properties of the log-growth rates (see the references above).
To the best of our knowledge, it has not been used before
in the study of the log-growth rates of city sizes.

The double mixture exponential GB2 (dmeGB2) distribution has
been introduced in \cite{Ram17} for the first time, and we will use it here with the same parametrization, and for the sake of brevity we refer to that article for its definition.
This distribution depends on 10 parameters
$(\rho,\epsilon,\nu,a,b,p,q,\tau,\zeta,\theta)$.

Mixtures of $m\geq 2$ non-standardized Student's $t$ distribution have the density
\begin{eqnarray*}
f_{m{\rm St}}(g;\mu_1,\sigma_1,\nu_1,
\ldots,\mu_m,\sigma_m,\nu_m,p_1,\ldots,p_{m-1})
&=&\sum_{j=1}^{m-1}p_jf_{{\rm St}}(g;\mu_j,\sigma_j,\nu_j) \nonumber\\
& &\quad +\left(1-\sum_{j=1}^{m-1}p_j\right)f_{{\rm St}}(g;\mu_m,\sigma_m,\nu_m)
\nonumber
\end{eqnarray*}
where $\mu_1,\ldots,\mu_m\in\mathbb{R}$, $\sigma_1,\ldots,\sigma_m>0$, $\nu_1,\ldots,\nu_m>0$, and $0\le p_1,\ldots,p_{m-1},p_1+\dots+p_{m-1}\le1$. In this paper, we will consider 2-mixtures ($m=2$) and 3-mixtures ($m=3$).

There is an issue with the estimation of these mixtures because in their general form maximum likelihood estimation (MLE) is numerically unstable. To make the estimation feasible we fix a priori the degrees of freedom $\nu_1,\nu_2$ for $m=2$ or $\nu_1,\nu_2,\nu_3$ for $m=3$. In particular, we estimate the remaining parameters for the three scenarios
\begin{eqnarray}
&&f_{2{\rm St}}(g;\mu_1,\sigma_1,4,\mu_2,\sigma_2,12,p_1),\nonumber\\
&&f_{2{\rm St}}(g;\mu_1,\sigma_1,4,\mu_2,\sigma_2,39,p_1),\nonumber\\
&&f_{3{\rm St}}(g;\mu_1,\sigma_1,4,\mu_2,\sigma_2,12,\mu_3,\sigma_3,39,p_1,p_2)\nonumber
\end{eqnarray}
and call them 2St12, 2St39 and 3St, respectively.
The intuition behind this is, e.g., for the 3-mixture, that a log-growth rate $g$ is drawn from a Student's $t$ distribution with a small degree of freedom with probability $p_1$, from a distribution with moderately high degree of freedom with probability $p_2$, and with a high degree of freedom with probability $1-p_1-p_2$.

Of course, the particular values of the $\nu$'s are arbitrary and other choices can yield to estimates with higher log-likelihoods. However, some decision has to be taken (in order to avoid numerical problems when estimating $\nu$'s) and we have experienced good results with these choices. Moreover, the Student's $t$ mixtures above are very competitive compared to the dmeGB2 of \cite{Ram17}, in many cases better.

\subsection{The estimation procedure}\label{estimproc}

In this paper we have estimated the parameters of all the distributions for the used samples by Maximum Likelihood Estimation (MLE), numerically using the command {\tt mle} of the {\sc MATLAB$^\circledR$} software package, on an equal footing for all the parameters. However, the Standard Errors (SE) of the ML estimators have been computed independently using the software package {\sc Mathematica}$^\circledR$ according to the indications of  \cite{McCVin03} and \cite{EfrHin78}.

\subsection{The information criteria}\label{infcrit}

We have considered in this paper three well-known information criteria very well adapted to the ML estimation, in order to select one model {}from the eight studied. They are:
\begin{itemize}
\item[-] The Akaike Information Criterion (AIC) \cite{Aka74,BurAnd02,BurAnd04}, defined as
    $$
    AIC=2k-2\ln L^*
    $$
    where $k$ is the number of parameters of the distribution and $\ln L^*$ is the corresponding (maximum) log-likelihood.
    The minimum value of AIC corresponds (asymptotically) to the minimum value of the Kullback--Leibler divergence, so a model with the lowest AIC is selected from among the competitors.
\item[-] The Bayesian or Schwarz Information Criterion (BIC) \cite{BurAnd02,BurAnd04,Sch78}, defined as
    $$
    BIC=k \ln(n)-2 \ln L^*
    $$
    where $k$ is the number of parameters of the distribution, $n$ the sample size and $\ln L^*$ is as before. The BIC penalizes more heavily the number of parameters used than does the AIC.
    The model with the lowest BIC is selected according to this criterion.
\item[-] The Hannan--Quinn Information Criterion (HQC) \cite{BurAnd02,BurAnd04,HanQui79}, defined as
    $$HQC=2 k \ln(\ln(n))-2 \ln L^*
    $$
    where $k$ is the number of parameters of the distribution, $n$ the sample size and $\ln L^*$ is as before. The HQC implements an intermediate penalization of the number of parameters when compared to the AIC and BIC. The model with the lowest HQC is selected according to this criterion.
\end{itemize}

\subsection{The log-rank/corank plots}\label{lrlcr}

We have employed in this article a standard technique for assessing the adequacy of an hypothesized distribution to the empirical data, as they are the log-rank/corank plots.
The main advantage of this approach is that if the tails are exponential, they should look like straight lines in the plots. This can be explained as follows. Assume that for the upper tail the probability density function for the log-growth rates $g$ takes the exponential form
$$
f_{{\rm eut}}(g;c_u)=c_u {\rm e}^{-c_u(g-g_m)}\,,\quad g\in[g_m,\infty)\,,
$$
where $g_m$ is a lower threshold for which this specification holds, and $c_u>0$.
Then the cumulative distribution function is, in the same support,
$$
{\rm cdf}_{{\rm eut}}(g;c_u)=1-{\rm e}^{-c_u(g-g_m)}
$$
The associated log-rank is $\ln[n_u(1-{\rm cdf}_{{\rm eut}}(g;c_u))]=\ln(n_u)-c_u(g-g_m)$, where $n_u$ is the sample size associated to the support where this law is hypothesized. The result is a straight line with negative slope. Analogously, we could consider the case where we have that for the lower tail it holds an exponential pdf
$$
f_{{\rm elt}}(g;c_l)=c_l {\rm e}^{c_l(g-g_M)}\,,\quad g\in(-\infty,g_M]\,,
$$
where $g_M$ is an upper threshold, and $c_l>0$.
Then the cumulative distribution function is, in the same support,
$$
{\rm cdf}_{{\rm elt}}(g;c_l)={\rm e}^{c_l(g-g_M)}
$$
The associated log-corank is $\ln[n_l {\rm cdf}_{{\rm elt}}(g;c_l)]=\ln(n_l)+c_l(g-g_M)$, where $n_l$ is the sample size associated to the support where this second law is assumed to hold. Thus for the log-corank, if the underlying law is (an increasing) exponential, we obtain a straight line with positive slope. Let us note that if we select $g_m=g_M$ in the mentioned two laws, the logarithm of the resulting global pdf is a tent-shaped function, as, e.g., in \cite{FuPamBulRicMatYamSta05}.

\section{The databases}\label{databases}

In this article we use population data, without
size restriction, of four European countries:
France, Germany, Italy and Spain \cite{PueRam15}, and
the samples described in \cite{Ram17} for USA cities' log-growth rates.

For the case of France, as in \cite{GonLanSan13},
we consider the lowest spatial subdivision, the \emph{communes},
as listed by the \emph{Institut National de la Statistique et des \'{E}tudes \'{E}conomiques} ({\tt www.insee.fr}). We have
employed the data for the years 1999 and 2009. For the two
samples we have aggregated the populations of the \emph{arrondisements} of the three biggest cities (which are also \emph{communes}): Paris, Marseille and Lyon.

For the case of Germany, Italy and Spain, the administrative urban unit of the data is the municipality (\emph{Gemeinden} for the case of Germany). For Germany, we take the data used in \cite{SchTre13} (the original source is the Federal German Statistical Office ({\tt www.destatis.de})). We take the data of the years 1996 and 2006 because for other years the frequent splittings and mergers of urban centres in Germany makes the computation of log-growth rates much more difficult.
For Italy, the data is obtained from the \emph{Istituto Nazionale di Statistica} ({\tt www.istat.it}), with all the Italian
municipalities (\emph{comuni}) for the years 2001 and 2011.
The data for Spain is taken from the \emph{Instituto Nacional de Estad\'{\i}stica} ({\tt www.ine.es}). They cover all the municipalities (\emph{municipios}) in the
years 2001 and 2010. See also \cite{GonLanSan14}

The data for the USA include log-growth rates of Incorporated Places (Ip) in the period 1990-2000, All Places (Ap) in the period 2000-2010, CCA clusters for the radius of 2km. in the period 1991-2000 \citep{RozRybGabMak11}, and that of Places that are one decade old in 1910 (d1 1910), five decades old in 1950 (d5 1950) and nine decades old in 1990 (d9 1990) \citep{SanGonVil14}. They are described thoroughly
in \cite{Ram17}. See Table~\ref{tgr} for the descriptive statistics of the employed log-growth data for France, Germany, Italy, Spain and the USA.

\begin{table}[htbp]
  \centering
  \caption{Descriptive statistics of the log-growth rates}
  \begin{footnotesize}
    \begin{tabular}{lccccc}
    \hline
     & Obs   & Mean  & SD    & Min   & Max \\
    France 1999--2009 & 36,643 & 0.099 & 0.150 & -2.060 & 2.692 \\
    Germany 1996--2006 & 12,309 & 0.007 & 0.112 & -0.827 & 1.006 \\
    Italy 2001--2011 & 8,081 & 0.043 & 0.117 & -0.580 & 3.303 \\
    Spain 2001--2010 & 8,074 & 0.038 & 0.244 & -1.458 & 3.258 \\
    USA Ip 1990--2000 & 19,048 & 0.075 & 0.262 & -4.467 & 3.581 \\
    USA Ap 2000--2010 & 24,685 & 0.035 & 0.282 & -5.278 & 6.075 \\
    USA CCA (2 km) 1991--2000 & 30,201 & 0.105 & 0.156 & -2.398 & 3.773 \\
    USA d1 1910 & 3291 & 0.186 & 0.415 & -1.914 & 3.723 \\
    USA d5 1950 & 3088 & 0.047 & 0.312 & -2.398 & 2.705 \\
    USA d9 1990 & 2987 & 0.056 & 0.261 & -1.580 & 3.581 \\
    \hline
    \end{tabular}%
    \end{footnotesize}
  \label{tgr}%
\end{table}%

\section{Results}\label{results}

As mentioned, we have estimated the studied distributions by
the method of Maximum Likelihood (ML),
using the command {\tt mle} of {\sc MATLAB$^\circledR$}.
We have reported in Tables~\ref{estimstu}, \ref{estimaln}, \ref{estimasb}, \ref{tedmesme}, \ref{2St12}, \ref{2St39} and \ref{3St} the estimated values
of the parameters for the Student's $t$, adLn, aSub,
dmeGB2, 2St12, 2St39 and 3St and the corresponding standard errors (SE)
computed according to \cite{EfrHin78} and \cite{McCVin03} with the software package {\sc Mathematica}$^\circledR$.
The ML estimators for the parameters of the normal
distribution are exact, being the mean and standard
deviation of each empirical data sample,
see Table~\ref{tgr}.
We can observe that the estimations are significative at the 5\% level in all cases but one, namely the $\mu$ parameter of Table~\ref{estimasb} for the aSub and the sample of USA d9 1990.

\begin{table}[htbp]
  \centering
  \caption{ML estimators and standard errors (SE) for the Student's $t$ and the European samples. Those for the USA samples are shown in Table~2 of \cite{Ram17}.}
  \begin{footnotesize}
      \begin{tabular}{llll}
    \hline
          & Student's $t$ &       &  \\
          & $\mu$ (SE) & $\sigma$ (SE) & $\nu$ (SE) \\
    France 1999--2009 & 0.092 (0.001) & 0.115 (0.001) & 5.236 (0.101)\\
    Germany 1996--2006 & 0.007 (0.001) & 0.084 (0.001) & 4.539 (0.137) \\
    Italy 2001--2011 & 0.040 (0.001) & 0.090 (0.001) & 5.715 (0.253) \\
    Spain 2001--2010 & 0.011 (0.002) & 0.166 (0.002) & 3.706 (0.120) \\
    \hline
    \end{tabular}%
    \end{footnotesize}
  \label{estimstu}
\end{table}%

\begin{table}[htbp]
  \centering
  \caption{ML estimators and standard errors (SE) for the asymmetric double Laplace normal (adLn) and the European samples. Those for the USA samples are shown in Table~3 of \cite{Ram17}.}
  \begin{footnotesize}
      \begin{tabular}{lllll}
    \hline
          & adLn &       &  & \\
          & $\alpha$ (SE) & $\beta$ (SE) & $\mu$ (SE) & $\sigma$ (SE) \\
     France 1999--2009 & 9.246 (0.058) & 14.187 (0.112) & 0.061 (0.001) & 0.066 (0.001)\\
    Germany 1996--2006 & 14.206 (0.165) & 14.346 (0.168) & 0.007 (0.001) & 0.047 (0.001) \\
     Italy 2001--2011 & 13.179 (0.188) & 17.294 (0.288) & 0.025 (0.001) & 0.057 (0.001) \\
     Spain 2001--2010 & 5.042 (0.062) & 10.556 (0.186) & -0.065 (0.002) & 0.081 (0.003) \\
    \hline
    \end{tabular}%
    \end{footnotesize}
  \label{estimaln}
\end{table}%

\begin{table}[htbp]
  \centering
  \caption{ML estimators and standard errors (SE) for the asymmetric Subbotin (aSub) and the studied log-growth rate samples. The estimate non-significative at the 5\% level is marked in bold.}
  \begin{footnotesize}
      \begin{tabular}{llllll}
    \hline
          & aSub &       &  & \\
          & $a_l$ (SE) & $a_r$ (SE) & $b_l$ (SE) & $b_r$ (SE) & $\mu$ (SE) \\
     France 1999--2009 &  0.0847 (0.0006) &  0.1491 (0.0007) & 1.072 (0.011) & 1.349 (0.011) & 0.0385 (0.0007) \\
    Germany 1996--2006 &  0.1043 (0.0009) & 0.0732 (0.0008) & 1.454 (0.022) & 0.957 (0.016) & 0.0375 (0.0009) \\
     Italy 2001--2011 & 0.110 (0.001) & 0.082 (0.001) & 1.802 (0.035) & 0.994 (0.020) & 0.069 (0.001) \\
     Spain 2001--2010 & 0.146 (0.002) & 0.209 (0.003) & 1.457 (0.032) & 1.014 (0.018) & -0.018 (0.002) \\
      USA Ip 1990-2000 & 0.103 (0.001) & 0.177 (0.001) & 0.767 (0.009) & 0.811 (0.008) & -0.008 (0.001) \\
    USA Ap 2000-2010 & 0.106 (0.001) & 0.147 (0.001) & 0.716 (0.006) & 0.669 (0.006) & -0.012 (0.001) \\
    USA CCA 1991-2000 (2km) & 0.052 (0.001) & 0.150 (0.001) & 0.889 (0.010) & 1.137 (0.009) & 0.008 (0.001) \\
    USA d1 1910 & 0.208 (0.005) & 0.356 (0.006) & 0.903 (0.029) & 1.053 (0.029) & 0.035 (0.005) \\
    USA d5 1950 & 0.180 (0.004) & 0.217 (0.005) & 1.094 (0.033) & 0.817 (0.023) & 0.014 (0.003) \\
    USA d9 1990 & 0.124 (0.003) & 0.177 (0.004) & 0.883 (0.027) & 0.834 (0.023) & \textbf{0.000 (0.003)} \\
    \hline
    \end{tabular}%
    \end{footnotesize}
  \label{estimasb}
\end{table}%

\begin{table}[htbp]
  \centering
  \caption{ML estimators and standard errors (SE) for the dmeGB2 and the European samples. Those for the USA samples are shown in Table~4 of \cite{Ram17}.}
  \begin{footnotesize}
      \begin{tabular}{lllll}
    \hline
          & dmeGB2 &       &  & \\
          & $\rho$ (SE) & $\epsilon$ (SE) & $\nu$ (SE) &  \\
    France 1999--2009 & 2.85 (0.29) & -0.162 (0.008) & 0.154 (0.017) & \\
     Germany 1996--2006 & 10.59 (0.18) & 0.031 (0.002) & 0.422 (0.019) & \\
    Italy 2001--2011 & 19.46 (0.80) & -0.106 (0.008) & 0.715 (0.092) & \\
    Spain 2001--2010 & 4.65 (1.41) & -0.604 (0.045) & 0.636 (0.220) & \\
          &       &       &  & \\
          & $a$ (SE) & $b$ (SE) & $p$ (SE) & $q$ (SE) \\
     France 1999--2009 & 13.07 (0.06) & 0.026 (0.001) & 1.454 (0.008) & 0.801 (0.004) \\
     Germany 1996--2006 & 5.86 (0.03) & -0.176 (0.001) & 13.866 (0.100) & 5.483 (0.030) \\
     Italy 2001--2011 & 21.79 (0.25) & 0.035 (0.001) & 0.659 (0.011) & 0.651 (0.008)\\
      Spain 2001--2010 & 12.73 (0.13) & -0.082 (0.002) & 0.910 (0.012) & 0.470 (0.006)\\
          &       &       &  & \\
          & $\tau$ (SE) & $\zeta$ (SE) & $\theta$ (SE) &  \\
    France 1999--2009 & -0.00 (0.06) & 2.04 (0.19) & 0.006 (0.001) &  \\
    Germany 1996--2006 & 0.17 (0.01) & 7.53 (0.40) & 0.56 (0.03) & \\
    Italy 2001--2011 & 0.35 (0.04) & 1.74 (0.65) & 0.13 (0.05) & \\
    Spain 2001--2010 & -0.14 (0.03) & 2.22 (0.14) & 0.07 (0.01) & \\
    \hline
    \end{tabular}%
    \end{footnotesize}
  \label{tedmesme}
\end{table}%

\begin{table}[htbp]
  \centering
  \caption{ML estimators and standard errors (SE) for the 2St12 and all samples. The parameters corresponding to the degrees of freedom are fixed a priori.}
  \begin{footnotesize}
    \begin{tabular}{lllll}
    \hline
          & 2St12 &       &       &  \\
          & $\nu_1$   & $\mu_1$ (SE) & $\sigma_1$ (SE) & $\nu_2$ \\
    France 1999-2009 & 4 & 0.175 (0.002) & 0.120 (0.001) & 12 \\
    Germany 1996-2006 & 4 & 0.021 (0.007) & 0.155 (0.005)  & 12 \\
    Italy 2001-2011 & 4 & 0.169 (0.013) & 0.138 (0.009) & 12 \\
    Spain 2001-2010 & 4 & 0.271 (0.010) & 0.240 (0.007) & 12 \\
    USA Ip 1990-2000 & 4 & 0.174 (0.005) & 0.273 (0.004) & 12 \\
    USA Ap 2000-2010 & 4 & 0.112 (0.004) & 0.284 (0.003) & 12 \\
    USA CCA 1991-2000 (2km) & 4 & 0.197 (0.002) & 0.123 (0.001) & 12 \\
    USA d1 1910 & 4 & 0.300 (0.014) & 0.398 (0.011) & 12 \\
    USA d5 1950 & 4 & 0.250 (0.023) & 0.387 (0.017) & 12 \\
    USA d9 1990 & 4 & 0.144 (0.015) & 0.299 (0.011) & 12 \\
          &       &       &       &  \\
          & $\mu_2$ (SE) & $\sigma_2$ (SE) & $p_1$ (SE) &  \\
    France 1999-2009 & 0.047 (0.001) & 0.090 (0.001) & 0.407 (0.005) &  \\
    Germany 1996-2006 & 0.005 (0.001) & 0.084 (0.001) & 0.126 (0.008) &  \\
    Italy 2001-2011 & 0.034 (0.001) & 0.091 (0.001) & 0.062 (0.006) &  \\
    Spain 2001-2010 & -0.024 (0.002) & 0.146 (0.002) & 0.198 (0.007) &  \\
    USA Ip 1990-2000 & 0.018 (0.001) & 0.102 (0.001) & 0.339 (0.006) &  \\
    USA Ap 2000-2010 & -0.010 (0.001) & 0.091 (0.001) & 0.337 (0.004) &  \\
    USA CCA 1991-2000 (2km) & 0.046 (0.001) & 0.070 (0.001) & 0.360 (0.004) &  \\
    USA d1 1910 & 0.078 (0.007) & 0.193 (0.006) & 0.466 (0.017) &  \\
    USA d5 1950 & -0.009 (0.004) & 0.169 (0.004) & 0.211 (0.013) &  \\
    USA d9 1990 & 0.015 (0.003) & 0.120 (0.003) & 0.285 (0.015) &  \\
    \hline
    \end{tabular}%
    \end{footnotesize}
  \label{2St12}%
\end{table}%

\begin{table}[htbp]
  \centering
  \caption{ML estimators and standard errors (SE) for the 2St39 and all samples. The parameters corresponding to the degrees of freedom are fixed a priori.}
  \begin{footnotesize}
    \begin{tabular}{lllll}
         \hline
          & 2St39 &       &       &  \\
          & $\nu_1$   & $\mu_1$ (SE) & $\sigma_1$ (SE) & $\nu_2$ \\
    France 1999-2009 & 4 & 0.164 (0.001) & 0.121 (0.001) & 39 \\
    Germany 1996-2006 & 4 & 0.014 (0.001) & 0.078 (0.001) & 39 \\
    Italy 2001-2011 & 4 & 0.077 (0.003) & 0.095 (0.002) & 39 \\
    Spain 2001-2010 & 4 & 0.248 (0.009) & 0.243 (0.007) & 39 \\
    USA Ip 1990-2000 & 4 & 0.157 (0.004) & 0.263 (0.003) & 39 \\
    USA Ap 2000-2010 & 4 & 0.100 (0.004) & 0.273 (0.003) & 39 \\
    USA CCA 1991-2000 (2km) & 4 & 0.201 (0.002) & 0.125 (0.001) & 39 \\
    USA d1 1910 & 4 & 0.284 (0.013) & 0.391 (0.010) & 39 \\
    USA d5 1950 & 4 & 0.197 (0.019) & 0.365 (0.014) & 39 \\
    USA d9 1990 & 4 & 0.117 (0.012) & 0.275 (0.009) & 39 \\
          &       &       &       &  \\
          & $\mu_2$ (SE) & $\sigma_2$ (SE) & $p_1$ (SE) &  \\
    France 1999-2009 & 0.043 (0.001) & 0.091 (0.001) & 0.464 (0.005) &  \\
    Germany 1996-2006 & -0.119 (0.004) & 0.037 (0.003) & 0.946 (0.004) &  \\
    Italy 2001-2011 & 0.014 (0.002) & 0.084 (0.001) & 0.447 (0.016) &  \\
    Spain 2001-2010 & -0.026 (0.002) & 0.149 (0.002) & 0.225 (0.008) &  \\
    USA Ip 1990-2000 & 0.017 (0.001) & 0.101 (0.001) & 0.380 (0.006) &  \\
    USA Ap 2000-2010 & -0.010 (0.001) & 0.091 (0.001) & 0.370 (0.005) &  \\
    USA CCA 1991-2000 (2km) & 0.047 (0.001) & 0.073 (0.001) & 0.350 (0.004) &  \\
    USA d1 1910 & 0.075 (0.007) & 0.191 (0.006) & 0.506 (0.017) &  \\
    USA d5 1950 & -0.012 (0.004) & 0.167 (0.004) & 0.268 (0.014) &  \\
    USA d9 1990 & 0.013 (0.004) & 0.116 (0.003) & 0.357 (0.016) &  \\
    \hline
    \end{tabular}%
    \end{footnotesize}
  \label{2St39}%
\end{table}%

\begin{table}[htbp]
  \centering
  \caption{ML estimators and standard errors (SE) for the 3St and all samples. The parameters corresponding to the degrees of freedom are fixed a priori.}
  \begin{footnotesize}
    \begin{tabular}{lllllll}
    \hline
          & 3St &       &       &       &       &  \\
          & $\nu_1$ & $\mu_1$ (SE) & $\sigma_1$ (SE) & $\nu_2$   & $\mu_2$ (SE) & $\sigma_2$ (SE) \\
    France 1999-2009 & 4     & 0.175 (0.002) & 0.120 (0.001) & 12 & 0.046 (0.001) & 0.091 (0.001) \\
    Germany 1996-2006 & 4     & 0.016 (0.002) & 0.107 (0.002) & 12 & 0.019 (0.001) & 0.065 (0.001) \\
    Italy 2001-2011 & 4     & 0.296 (0.031) & 0.172 (0.023) & 12 & 0.044 (0.002) & 0.104 (0.001) \\
    Spain 2001-2010 & 4     & 0.373 (0.015) & 0.272 (0.012) & 12 & 0.012 (0.003) & 0.168 (0.002) \\
    USA Ip 1990-2000 & 4     & 0.292 (0.012) & 0.384 (0.008) & 12 & 0.065 (0.003) & 0.175 (0.002) \\
    USA Ap 2000-2010 & 4     & 0.238 (0.012) & 0.452 (0.009) & 12 & 0.032 (0.002) & 0.185 (0.002) \\
    USA CCA 1991-2000 (2km) & 4 & 0.280 (0.005) & 0.181 (0.003) & 12 & 0.139 (0.001) & 0.093 (0.001) \\
    USA d1 1910 & 4     & 0.045 (0.008) & 0.173 (0.007) & 12 & 0.258 (0.013) & 0.274 (0.011) \\
    USA d5 1950 & 4 & 0.369 (0.042) & 0.482 (0.030) & 12 & -0.014 (0.005) & 0.143 (0.004) \\
    USA d9 1990 & 4 & 0.269 (0.038) & 0.429 (0.027) & 12 & 0.014 (0.004) & 0.105 (0.003) \\
          &       &       &       &       &       &  \\
          & $\nu_3$ & $\mu_3$ (SE) & $\sigma_3$ (SE) & $p_1$ (SE) & $p_2$ (SE) &  \\
    France 1999-2009 & 39    & 0.047 (0.010) & 0.012 (0.006) & 0.408 (0.002) & 0.588 (0.002) &  \\
    Germany 1996-2006 & 39    & -0.109 (0.003) & 0.046 (0.003) & 0.398 (0.005) & 0.515 (0.005) &  \\
    Italy 2001-2011 & 39    & 0.026 (0.003) & 0.071 (0.003) & 0.015 (0.002) & 0.706 (0.022) &  \\
    Spain 2001-2010 & 39    & -0.094 (0.007) & 0.087 (0.006) & 0.109 (0.005) & 0.752 (0.012) &  \\
    USA Ip 1990-2000 & 39    & 0.008 (0.002) & 0.080 (0.001) & 0.139 (0.004) & 0.482 (0.008) &  \\
    USA Ap 2000-2010 & 39    & -0.013 (0.001) & 0.074 (0.001) & 0.118 (0.003) & 0.439 (0.006) &  \\
    USA CCA 1991-2000 (2km) & 39    & 0.027 (0.001) & 0.063 (0.001) & 0.126 (0.003) & 0.405 (0.005) &  \\
    USA d1 1910 & 39    & 0.393 (0.038) & 0.726 (0.027) & 0.446 (0.011) & 0.389 (0.013) &  \\
    USA d5 1950 & 39    & 0.042 (0.012) & 0.268 (0.009) & 0.110 (0.010) & 0.541 (0.026) &  \\
    USA d9 1990 & 39    & 0.054 (0.010) & 0.224 (0.008) & 0.108 (0.011) & 0.545 (0.023) &  \\
    \hline
    \end{tabular}%
    \end{footnotesize}
  \label{3St}%
\end{table}%

\begin{table}
  \caption{Maximum log-likelihoods, AIC, BIC and HQC for the distributions and samples. The lowest values of AIC, BIC and HQC for each sample are marked in bold.}
  \begin{footnotesize}
       \begin{tabular}{lllllllll}
       \hline
          & normal &       &       &       & Student's $t$ &       &       &  \\
          & log-likelihood & AIC & BIC & HQC & log-likelihood & AIC & BIC & HQC \\
    France 1999-2009 & 17477 & -34951 & -34934 & -34945 & 20122 & -40239 & -40213 & -40230 \\
    Germany 1996-2006 & 9432  & -18859 & -18845 & -18854 & 10202 & -20398 & -20376 & -20391 \\
    Italy 2001-2011 & 5855  & -11705 & -11691 & -11701 & 65413 & -13020 & -12999 & -13013 \\
    Spain 2001-2010 & -80   & 165   & 179   & 169   & 749   & -1492 & -1471 & -1485 \\
    USA Ip 1990-2000 & -1548 & 3100  & 3116  & 3106  & 3067  & -6129 & -6105 & -6121 \\
    USA Ap 2000-2010 & -3817 & 7638  & 7655  & 7644  & 5244  & -10482 & -10458 & -10475 \\
    USA CCA 1991-2000 (2km) & 13302 & -26600 & -26584 & -26595 & 18284 & -36561 & -36536 & -36553 \\
    USA d1 1910 & -1775 & 3554  & 3567  & 3559  & -1443 & 2893  & 2911  & 2899 \\
    USA d5 1950 & -784  & 1572  & 1584  & 1576  & -301  & 608   & 626   & 615 \\
    USA d9 1990 & -228  & 459   & 471   & 463   & 410   & -813  & -795  & -807 \\
          &       &       &       &       &       &       &       &  \\
          & adLN &       &       &       & aSub &       &       &  \\
          & log-likelihood & AIC & BIC & HQC & log-likelihood & AIC & BIC & HQC \\
    France 1999-2009 & 20226 & -40443 & -40409 & -40433 & 20000 & -39990 & -39947 & -39976 \\
    Germany 1996-2006 & 10153 & -20298 & -20268 & -20288 & 10123 & -20236 & -20199 & -20224 \\
    Italy 2001-2011 & 6499  & -12990 & -12962 & -12980 & 6475  & -12941 & -12906 & -12929 \\
    Spain 2001-2010 & 925   & -1842 & -1814 & -1832 & 892   & -1775 & -1740 & -1763 \\
    USA Ip 1990-2000 & 3053  & -6098 & -6067 & -6088 & 3275  & -6540 & -6501 & -6527 \\
    USA Ap 2000-2010 & -- & -- & -- & -- & 5212  & -10414 & -10373 & -10401 \\
    USA CCA 1991-2000 (2km) & 19251 & -38493 & -38460 & -38483 & 19089 & -38169 & -38127 & -38155 \\
    USA d1 1910 & -1405 & 2818  & 2842  & 2827  & -1406 & 2823  & 2853  & 2833 \\
    USA d5 1950 & -295  & 598   & 622   & 606   & -295  & 600   & 630   & 610 \\
    USA d9 1990 & 384   & -760  & -736  & -752  & 398   & -785  & -755  & -774 \\
          &       &       &       &       &       &       &       &  \\
          & dmeGB2 &       &       &       & 2St12 &       &       &  \\
          & log-likelihood & AIC & BIC & HQC & log-likelihood & AIC & BIC & HQC \\
    France 1999-2009 & 20511 & -41003 & -40918 & -40976 & 20513 & \textbf{-41016} & \textbf{-40973} & \textbf{-41002} \\
    Germany 1996-2006 & 10242 & \textbf{-20463} & -20389 & -20438 & 10213 & -20415 & -20378 & -20403 \\
    Italy 2001-2011 & 6534  & -13049 & -12979 & -13025 & 6534  & \textbf{-13057} & \textbf{-13022} & \textbf{-13045} \\
    Spain 2001-2010 & 954   & -1887 & -1817 & -1863 & 941   & -1873 & -1838 & -1861 \\
    USA Ip 1990-2000 & 3509  & -6998 & -6920 & -6972 & 3466  & -6921 & -6882 & -6909 \\
    USA Ap 2000-2010 & 5625  & -11231 & -11150 & -11204 & 5553  & -11097 & -11056 & -11084 \\
    USA CCA 1991-2000 (2km) & 19771 & \textbf{-39521} & \textbf{-39438} & \textbf{-39495} & 19578 & -39146 & -39105 & -39133 \\
    USA d1 1910 & -1388 & \textbf{2796} & 2857  & 2817  & -1395 & 2800  & \textbf{2830} & \textbf{2810} \\
    USA d5 1950 & -254  & 529   & 589   & 550   & -258  & 527   & \textbf{557} & \textbf{537} \\
    USA d9 1990 & 446   & \textbf{-873} & -813  & -851  & 438   & -865  & \textbf{-835} & \textbf{-854} \\
          &       &       &       &       &       &       &       &  \\
          & 2St39 &       &       &       & 3St &       &       &  \\
          & log-likelihood & AIC & BIC & HQC & log-likelihood & AIC & BIC & HQC \\
    France 1999-2009 & 20502 & -40994 & -40952 & -40981 & 20514 & -41011 & -40943 & -40990 \\
    Germany 1996-2006 & 10237 & -20463 & \textbf{-20426} & \textbf{-20451} & 10241 & -20457 & -20398 & -20437 \\
    Italy 2001-2011 & 6527  & -13044 & -13009 & -13032 & 6535  & -13055 & -12999 & -13036 \\
    Spain 2001-2010 & 942   & -1874 & \textbf{-1839} & -1862 & 953   & \textbf{-1889} & -1833 & \textbf{-1870} \\
    USA Ip 1990-2000 & 3452  & -6894 & -6855 & -6881 & 3519  & \textbf{-7021} & \textbf{-6959} & \textbf{-7001} \\
    USA Ap 2000-2010 & 5526  & -11042 & -11002 & -11029 & 5641  & \textbf{-11265} & \textbf{-11200} & \textbf{-11244} \\
    USA CCA 1991-2000 (2km) & 19627 & -39244 & -39202 & -39230 & 19757 & -39497 & -39431 & -39476 \\
    USA d1 1910 & -1395 & 2800  & 2830  & 2811  & -1390 & 2796  & 2844  & 2813 \\
    USA d5 1950 & -261  & 533   & 563   & 543   & -255  & \textbf{526} & 575   & 544 \\
    USA d9 1990 & 435   & -861  & -831  & -850  & 443   & -870  & -822  & -853 \\
    \hline
    \end{tabular}%
    \end{footnotesize}
  \label{llAICBICHQC}%
\end{table}%

With these estimations, we have computed the Akaike Information Criterion (AIC), the Ba\-ye\-sian or Schwarz Information Criterion (BIC) and the  Hannan--Quinn Information Criterion (HQC) \cite{BurAnd02,BurAnd04}, see Table~\ref{llAICBICHQC}.
We note that the normal distribution is never chosen for any sample, despite of the fact of being the distribution with less parameters.

The selected models are among the dmeGB2 (only clearly for CCA clusters), 2St12, 2St39 and 3St, in a not very regular fashion. That is, they are all very competitive models and none dominates clearly over the others. These are the best models for log-growth rates of city sizes known to date, to the best of our knowledge, and
one cannot discard any of them in principle.

\begin{figure}
\begin{tabular}{ccc}
\epsfig{file=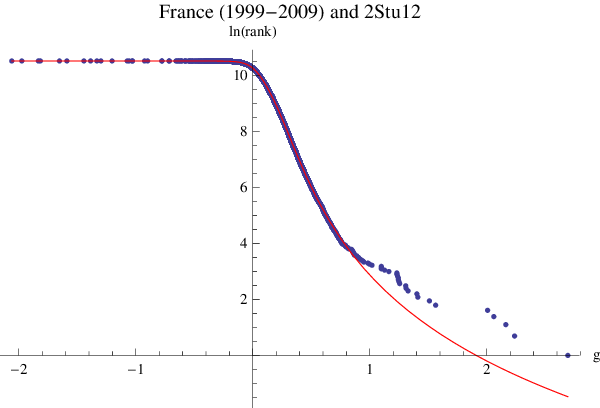, width=4cm} &
\epsfig{file=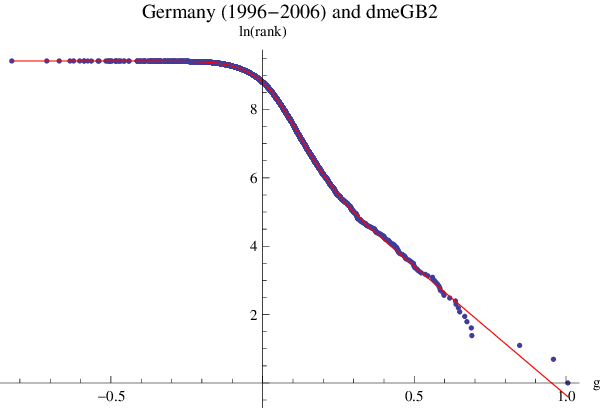, width=4cm}&
\epsfig{file=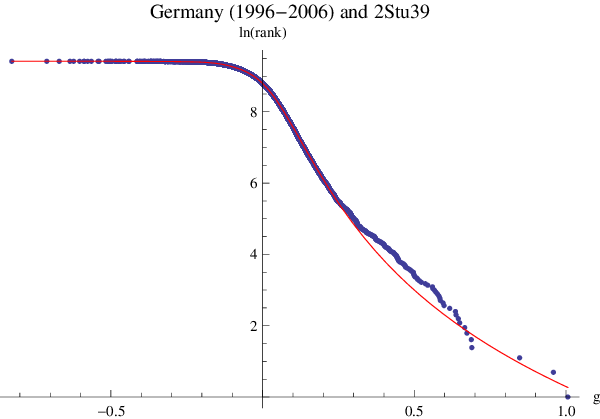, width=4cm}\\
\epsfig{file=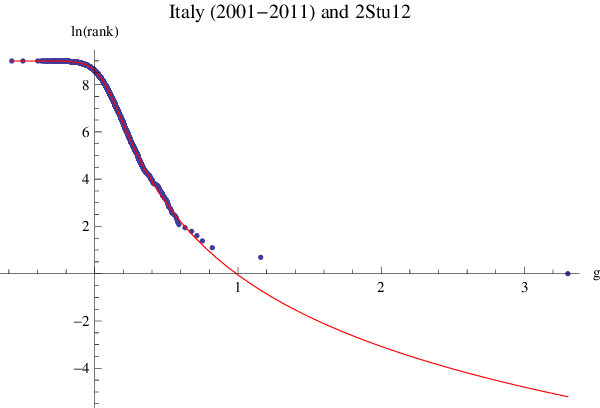, width=4cm} &
\epsfig{file=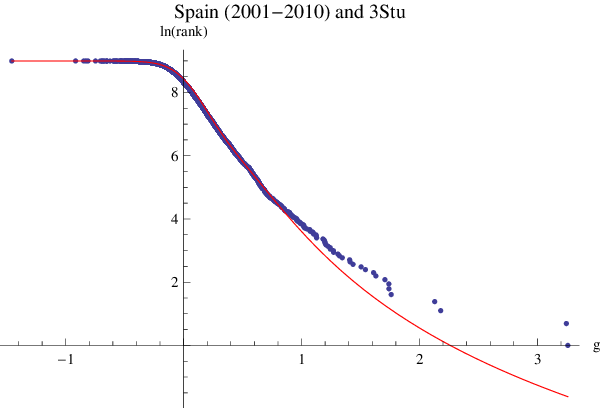, width=4cm}&
\epsfig{file=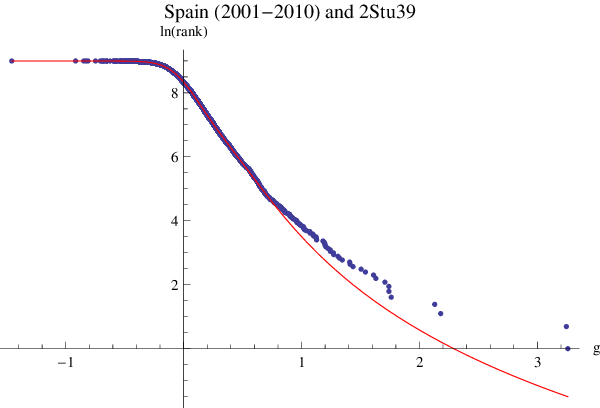, width=4cm}\\
\epsfig{file=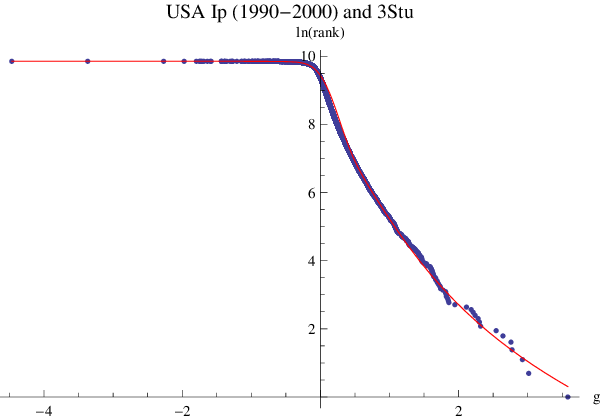, width=4cm} &
\epsfig{file=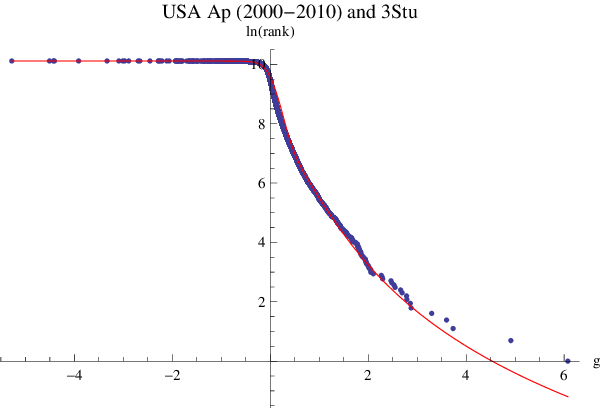, width=4cm}&
\epsfig{file=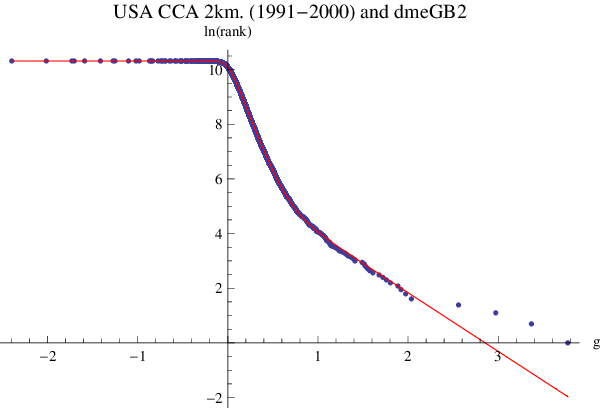, width=4cm}\\
\epsfig{file=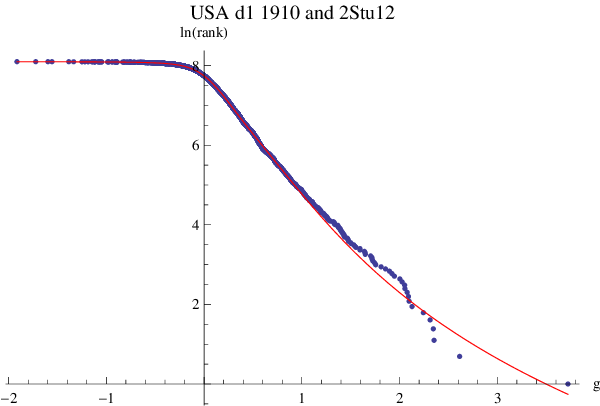, width=4cm} &
\epsfig{file=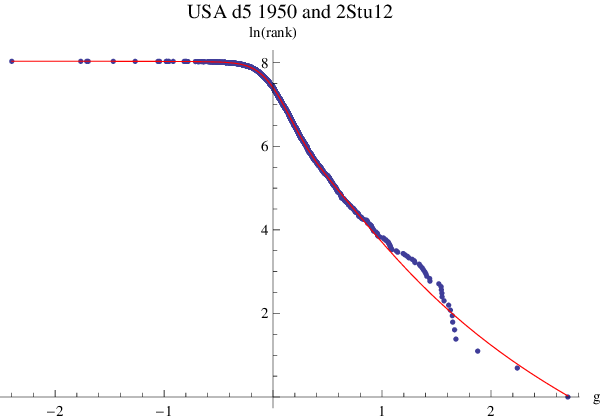, width=4cm}&
\epsfig{file=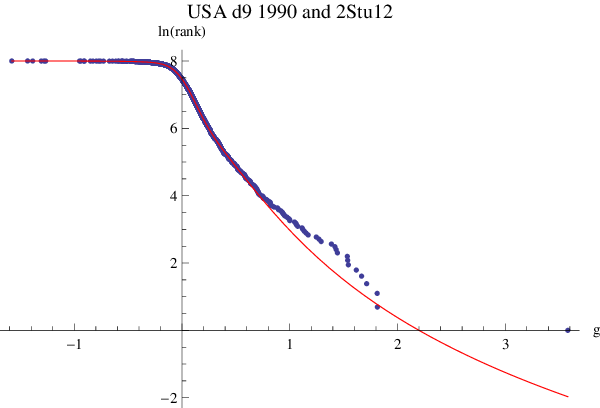, width=4cm}\\
\end{tabular}
\caption{Log-rank plots for the selected models by AIC, BIC, and HQC. Empirical points in blue, predicted model in red. (Color online).}
\label{figr}
\end{figure}

\begin{figure}
\begin{tabular}{ccc}
\epsfig{file=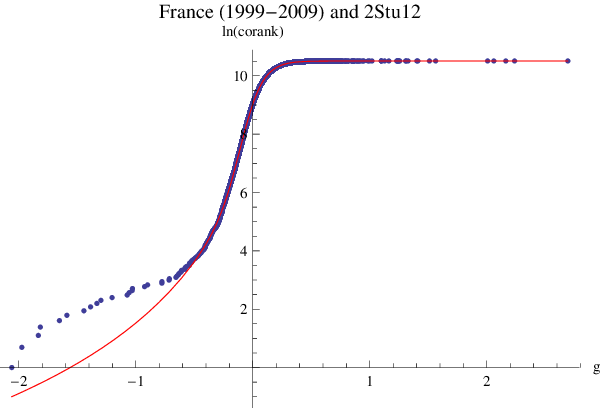, width=4cm} &
\epsfig{file=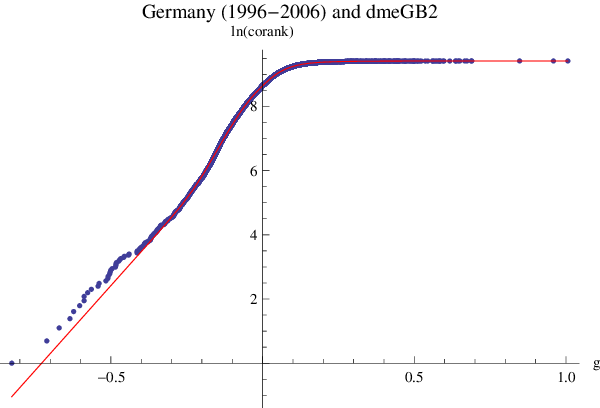, width=4cm}&
\epsfig{file=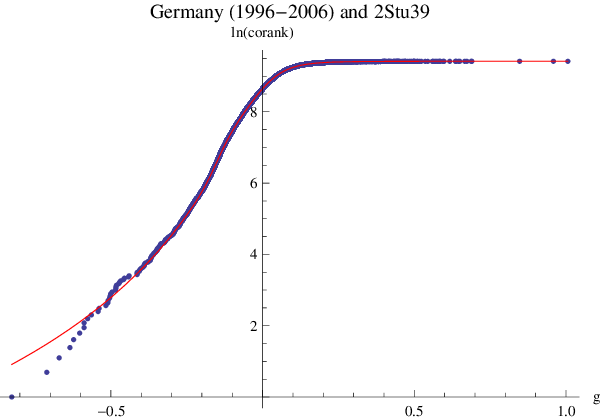, width=4cm}\\
\epsfig{file=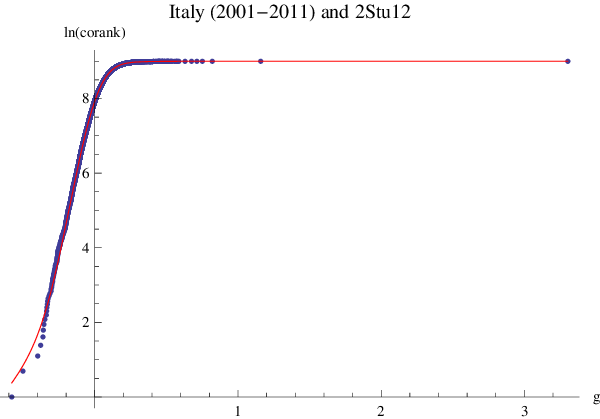, width=4cm} &
\epsfig{file=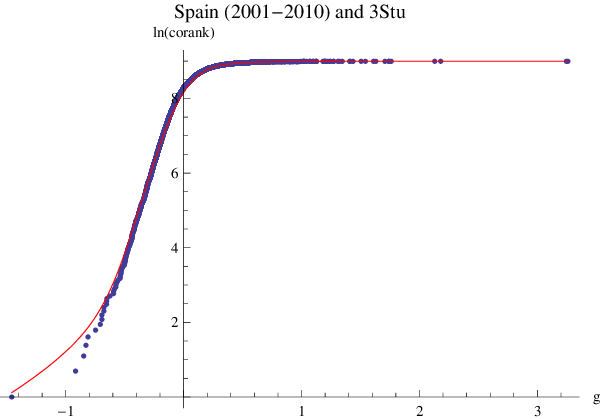, width=4cm}&
\epsfig{file=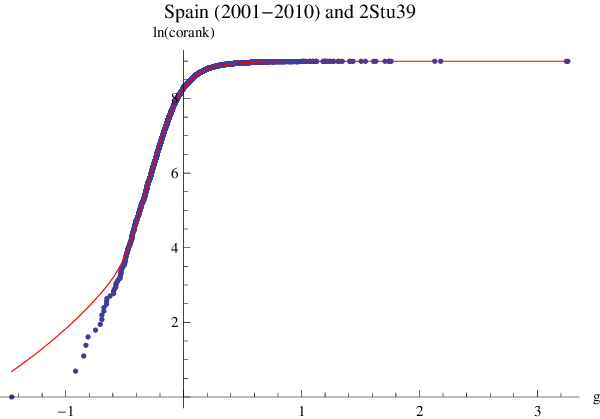, width=4cm}\\
\epsfig{file=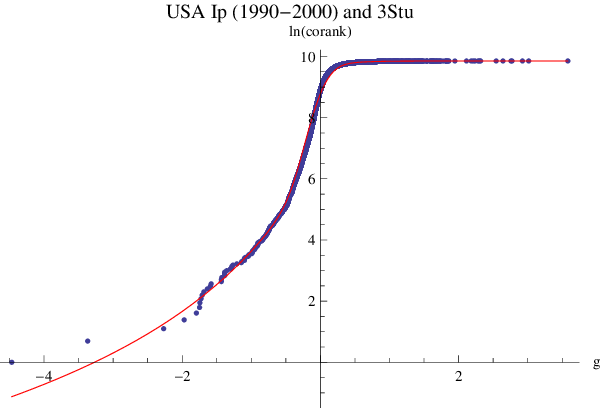, width=4cm} &
\epsfig{file=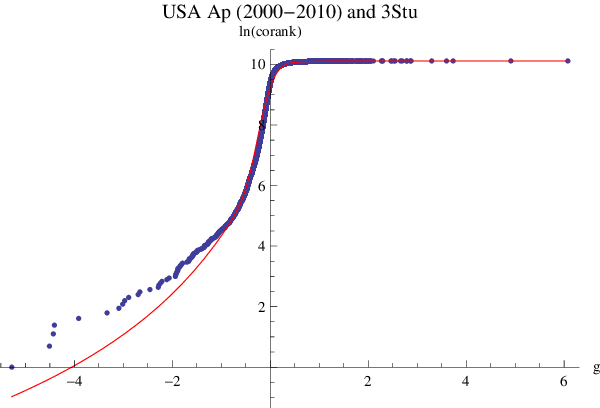, width=4cm}&
\epsfig{file=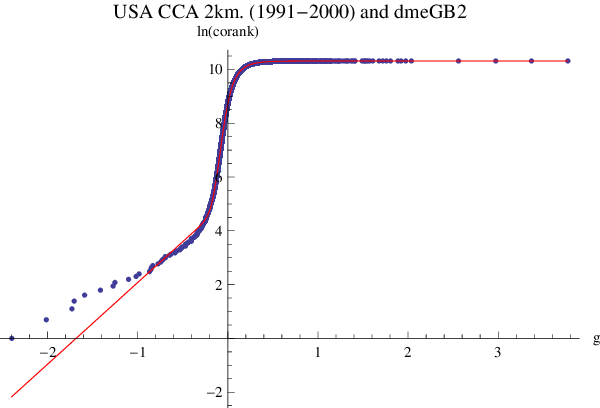, width=4cm}\\
\epsfig{file=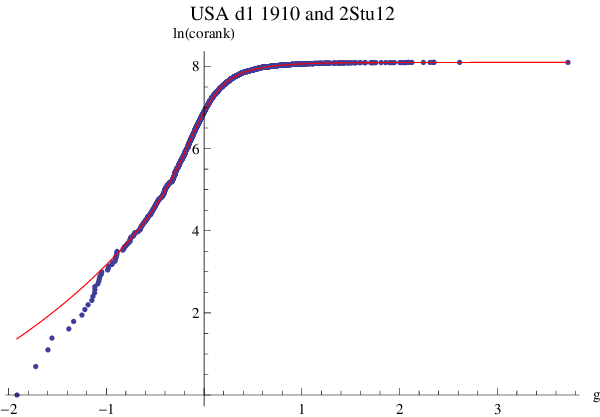, width=4cm} &
\epsfig{file=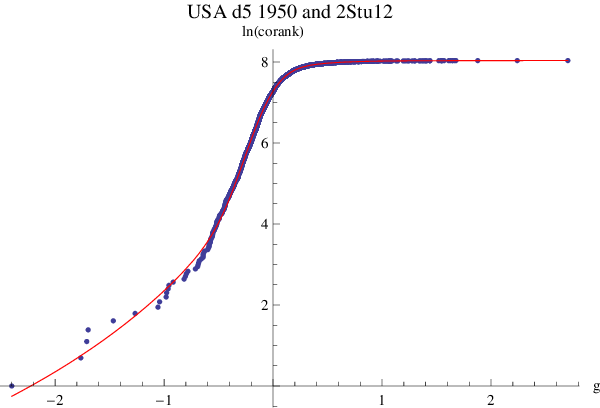, width=4cm}&
\epsfig{file=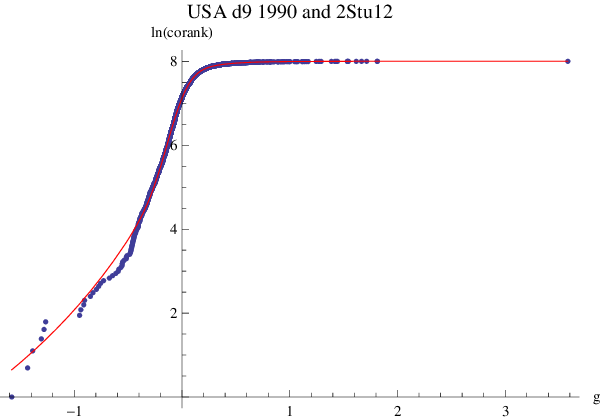, width=4cm}\\
\end{tabular}
\caption{Log-corank plots for the selected models by AIC, BIC, and HQC. Empirical points in blue, predicted model in red. (Color online).}
\label{figcr}
\end{figure}

In order to assess the goodness-of-fit,
we have used the standard log-rank/corank plots.
We have plotted these quantities for the selected models by AIC, BIC, and HQC for each sample, with the exception of the samples of USA d1 1910, d5 1950 and d9 1990 for which we have chosen the plots indicated by both BIC and HQC, for the sake of brevity.
The deviations at the upper (resp. lower) tails are
amplified by the fact of taking logarithms (see, e.g., \cite{GonRamSan13}). And if the tails are exponential, the plot at the upper/lower tails should be linear.
What we see at Figures~\ref{figr} and~\ref{figcr} is that the discrepancies are not very remarkable but that the graphs are mainly curved, as a difference with reported properties of firm growth rates \citep{bottazzi2006edf,BotSec11,CanAmaLeeMeySta98,Wit05,
GatGuiGafGiuGalPal05,
FagNapRov08,
FuPamBulRicMatYamSta05,
FujGuiAoyGalSou04,GalPal10,RicPamBulPonSta08,
Rozenfeld12022008,
StaAmaBulHavLesMaaSalSta96}.

\section{Conclusions}\label{discussionconcl}

We have seen in the preceding section that the normal
distribution is not a good description of the log-growth
rates of city sizes in four major European countries, and for the USA samples in \cite{Ram17} as well,
or at least it has been surpassed by a number of
alternative models. This renders questionable the standard
form of Gibrat's law for European and USA city sizes. This has been known for firm sizes and for the output of different countries for some time \citep{bottazzi2006edf,BotSec11,CanAmaLeeMeySta98,GatGuiGafGiuGalPal05,
FagNapRov08,
FuPamBulRicMatYamSta05,FujGuiAoyGalSou04,
RicPamBulPonSta08,
Rozenfeld12022008,
StaAmaBulHavLesMaaSalSta96}. According to these references, we have studied as well as a candidate the asymmetric Subbotin (aSub) distribution for the first time in the topic of log-growth rates of city sizes. This distribution performs better than the normal but is still surpassed by other models. We have added the asymmetric double Laplace normal (adLn) of \cite{Ree02,Ree03,reed2004double} to the study and it performs slightly better than the aSub. But we have seen that very appropriate parametric models for the log-growth rate distribution of the city sizes of France, Germany, Italy, Spain and the USA are the 2St12, 2St39, 3St mixtures of Student's $t$ and the dmeGB2, introduced in \cite{Ram17}.
It is also interesting the fact that the 2St12, 2St39 and 3St play a relevant role in the parametric description of stock market indices around the globe.\footnote{Work in preparation.}

This means that, in spite of the fact that the studied European countries and the USA have historically very different urban systems, with very different age distributions of urban centres
\citep{Cub11,GieSue14,SanGonVil14}, the distribution of log-growth rates can be described with essentially the same parametric models. From the theoretical side, the results might shed new light on the interpretation of current theories of urban growth. Indeed, knowing good fitting model distributions will eventually be crucial for making future predictions with appropriate determinants. This is  economically important for policy makers because \lq\lq extremely large investments in building new housing and infrastructure must be made to accommodate the demographic growth of cities.\rq\rq\ \citep{DurPug14}.

As a final remark to this study, we have seen that we have several different models that compete tightly among them, and one is slightly better than the others irrespectively of the fact that they correspond to the studied European or USA data sets, or the age of cities. There seems to be no regularity associated to these characteristics of the data, so there may seem to exist some deeper generation processes for them. 
This might be the case only for developed countries.
The comparison with what happens for developing countries 
should be the subject or another study, and for this task
it might be a difficulty the availability of good data sets.
 
\section*{Author contributions}
Till Massing: Conceptualization, formal analysis, investigation, methodology, software, supervision, validation, visualization, writing-review \& editing. Miguel Puente-Ajov\'{\i}n: Conceptualization, data curation, formal analysis, funding acquisition, investigation, methodology, resources, software, validation, visualization, writing-original draft, writing-review \& editing. Arturo Ramos: Conceptualization, data curation, formal analysis, funding acquisition, investigation, methodology, resources, software, validation, visualization, writing-original draft, writing-review \& editing.

\section*{Competing interests statement}
The authors declare to have no competing interests concerning the research carried out in this article.

\section*{Acknowledgments}
We would like to thank Rafael Gonz\'alez-Val, Fernando Sanz-Gracia and Mark Trede for the databases used.
The authors would like to thank the Editor and the two reviewers for a careful reading and comments which greatly improved the paper. The work of Miguel Puente-Ajov\'{\i}n and Arturo Ramos has been supported by the Spanish \emph{Ministerio de Econom\'{\i}a y Competitividad} (ECO2017-82246-P) and by Aragon Government (ADETRE Reference Group).

\bibliographystyle{apalike}
\bibliography{biblio}

\vfill\eject

\vfill\eject

\end{document}